\documentclass[reprint,aps,prb,floatfix,twocolumn,superscriptaddress,showpacs,amsmath,amssymb,showpacs,]{revtex4-1} 

\usepackage{amsmath,amssymb,graphics,epsfig,epstopdf,color,verbatim,ulem,braket,tabularx}
\usepackage{multirow}
\usepackage[colorlinks,linkcolor=blue,citecolor=blue,urlcolor=blue,bookmarks=false]{hyperref}

\usepackage{listings}
\usepackage{cancel}
\usepackage{mathrsfs}
\usepackage{soul}
\usepackage{color}
\usepackage{url}
\usepackage{siunitx}
\usepackage{float}
\usepackage{lipsum}
\usepackage{wrapfig}

\usepackage[dvipsnames]{xcolor}

\usepackage{booktabs}

\begin{document}

\title{Plasmon assisted superconductivity in LiTi$_2$O$_4$}

\author{Francesco Petocchi}
\email{francesco.petocchi@unifr.ch}
\affiliation{Department of Physics, University of Fribourg, 1700 Fribourg, Switzerland}

\author{Viktor Christiansson}
\email{viktor.christiansson@tuwien.ac.at}
\affiliation{Institute of Solid State Physics, TU Wien, 1040 Vienna, Austria}
\affiliation{Department of Physics, University of Fribourg, 1700 Fribourg, Switzerland}

\author{Ryotaro Arita}
\email{arita@riken.jp}
\affiliation{Department of Physics, The University of Tokyo, Hongo, Bunkyo-ku, Tokyo 113-0033, Japan}
\affiliation{RIKEN Center for Emergent Matter Science, 2-1 Hirosawa, Wako, Saitama 351-0198, Japan}

\author{Philipp Werner$^\dagger$}
\email{philipp.werner@unifr.ch}
\affiliation{Department of Physics, University of Fribourg, 1700 Fribourg, Switzerland}

\begin{abstract}
We combine $GW$ plus extended dynamical mean field theory ($GW$+EDMFT) with the density functional theory for superconductors (SCDFT) framework to study the electronic properties of LiTi$_2$O$_4$. Excellent agreement with experiment is obtained for the density of states, mass enhancement, Sommerfeld coefficient and 
superconducting $T_c$, if the dynamical nature of the screened Coulomb interaction is taken into account. Our results show that the coupling to collective charge fluctuations (plasmons) plays an important role in the pairing mechanism and explains the remarkably high $T_c$ of this moderately correlated spinel compound. 
\end{abstract}

\maketitle

{\it Introduction --} More than 200 compounds are known to crystallize in the AB$_2$O$_4$ normal spinel structure, but only a few of them exhibit superconductivity. Examples include 
CuRh$_2$Se$_4$
(with a critical temperature $T_c = 3.5$~K), CuV$_2$S$_4$ ($T_c = 4.5$~K) and CuRh$_2$S$_4$ ($T_c = 4.7$~K). Within this family, LiTi$_2$O$_4$ (LTO) is the only oxide superconductor and stands out with the highest critical temperature, ranging from 10~K to 13.7~K.\cite{Johnston1973,Heintz1989,Moshopoulou1999,Geng2005,Tang2006,Sun2004}

LTO is a type-II superconductor with a fully gapped $s$-wave order parameter, whose properties can be described to a good extent within the Bardeen–Cooper–Schrieffer (BCS) framework, where pairing is mediated by electron–phonon ($e$–$ph$) interactions.\cite{Heintz1989,Geng2005} However, comparisons between band-structure calculations, angle-resolved photoemission spectroscopy (ARPES) data and specific-heat measurements indicate significant electron–electron ($e$–$e$) correlations.\cite{Edwards1984,Johnston1999,Sun2004,Tang2006,hasan2025} In the normal state, the resistivity shows a metallic behavior, decreasing monotonically with temperature until the superconducting transition. The superconducting properties are highly sensitive to stoichiometry\cite{Fazileh2004,Yoshimatsu2015} and recent ionic liquid gating experiments\cite{Maruyama2015,Wei2021} reported a superconducting dome between two insulating states with unclear origin. Furthermore, LTO displays an angle-dependent anomalous magnetoresistance,\cite{Jin2015} which may originate from spin–orbital fluctuations reminiscent of those observed in high-$T_c$ cuprate superconductors.\cite{Xue2022}

Experimental investigations of LiTi$_2$O$_4$ reported values for both the specific-heat (Sommerfeld) coefficient $\gamma \approx 19$ mJ mol$^{-1}$ K$^{-2}$ and the superconducting gap amplitude $\Delta=1.97$ meV.\cite{Johnston1973,Ekino1990,Sun2004} Concurrently, the first density functional theory (DFT) studies\cite{Satpathy1987,Massidda1988} provided an estimate for the electronic density of states (DoS) at the Fermi level, $N_{\varepsilon_F}^{\text{DFT}} = 0.46$ states/eV~atom. By comparing this value with the one determined from $\gamma=\frac{K_B^{2}\pi^{2}}{3}N_{\varepsilon_F}^{\text{exp}}$, an $e$–$ph$ coupling constant $\lambda_{ph} \approx 1.8$ was initially inferred using the formula $N_{\varepsilon_F}^{\text{exp}} = (1 + \lambda_{ph}) N_{\varepsilon_F}^{\text{DFT}}$. However, even at this early stage, estimates of $T_c$ based on the McMillan formula \cite{McMillan1968,Allen1975} indicated that such a strong coupling was inconsistent with the experimentally observed critical temperature, which instead suggested an intermediate value of $\lambda_{ph} \approx 0.6$. This discrepancy provided the first indication that an additional renormalization mechanism, associated with $e$–$e$ interactions, was required to understand the properties of LTO. Consequently, the total enhancement of the DoS was reformulated as $N_{\varepsilon_F}^{\text{exp}} = (1 + \lambda_{ph} + \lambda_{ee}) N_{\varepsilon_F}^{\text{DFT}}$, implying $\lambda_{ph} + \lambda_{ee} \approx 1.8$. More recent experimental work\cite{Sun2004} refined these estimates, confirming $\lambda_{ph} \approx 0.65$ from both the specific-heat jump and independent fits using the McMillan equation. The same study also reported an updated specific-heat coefficient of $\gamma = 19.15$~mJ~mol$^{-1}$~K$^{-2}$. Combining this value with $N_{\varepsilon_F}^{\text{DFT}}$ and $\lambda_{ph} = 0.65$ yields $\lambda_{ph} + \lambda_{ee} \approx 1.53$, corresponding to $\lambda_{ee} \approx 0.88$.

These results imply that LiTi$_2$O$_4$ cannot be described as a purely $e$–$ph$ driven superconductor. In particular, the significant enhancement of the experimentally derived density of states, relative to the DFT prediction, suggests the presence of non-negligible $e$–$e$ interactions. The influence of $e$–$e$ correlations can be quantified by the quasiparticle renormalization factor $Z$, which based on the DFT and experimental results is $Z \approx 0.65$, a value consistent with moderate $e$–$e$ correlations. 

The recent ionic-gating study on LTO\cite{Wei2021} provides further evidence for nontrivial correlation effects. The experiment probed the superconducting response as a function of carrier concentration, effectively tuning the titanium valence between the limiting configurations $d^0$ and $d^1$. The $d^0$ case is rather straightforward, given that the oxygen $p$ bands lie approximately 2\,eV below the Ti $t_{2g}$ manifold and the resulting state is expected to be band-insulating. In contrast, the emergence of an insulating or bad metal phase in the $d^1$ limit cannot be explained within a simple single-particle picture, pointing instead to a correlation effect. 
Here, we will show that these correlations have important implications both on the doping evolution of the electronic structure and on the nature of the pairing glue in LTO, and more specifically that plasmons (collective charge fluctuations) are essential for explaing the high $T_c$ of this spinel compound. 

{\it Electronic structure --} The normal spinel AB$_2$O$_4$ adopts the cubic $\textit{Fd-3m}$ space group (No. 227), corresponding to a face-centered cubic (fcc) lattice. In LiTi$_2$O$_4$, Li$^+$ ions occupy the tetrahedral (8a) “A” sites, while Ti cations reside on the octahedral (16d) “B” sites, forming TiO$_6$ octahedra within an oxygen-based cubic close-packed framework with a lattice constant of approximately 8.37 Å.\cite{Campa1994,Takahashi2002} The Ti ions exhibit a mixed-valence configuration with equal proportions of Ti$^{3+}$ ($3d^1$) and Ti$^{4+}$ ($3d^0$), resulting in partially filled Ti $3d^{0.5}$ bands.\cite{Chen2011} The octahedral crystal field splits the Ti $3d$ manifold, pushing the $e_g$ states about 2~eV above the Fermi level, while the $t_{2g}$ orbitals form the low-energy manifold responsible for the metallic character of the compound.\cite{Chen2011,Liu2017}

Our electronic structure analysis is based on the real-space extension of $GW$+extended dynamical mean field theory ($GW$+EDMFT), which has previously been applied to the related material Ca$_2$RuO$_4$, \cite{Petocchi2021} where octahedral rotations make the local Hamiltonians for the sites within the unit cell inequivalent. The $GW$+EDMFT framework \cite{Biermann2003,Boehnke2016} captures screening, nonlocal exchange–correlations, and local dynamical fluctuations without resorting to empirical interaction parameters or heuristic double-counting corrections. Its fully diagrammatic formulation enables partitioning of the self-energy into high-, \mbox{intermediate-,} and low-energy contributions, which are treated, respectively, by single-shot $G^0W^0$, self-consistent $GW$, and EDMFT.\cite{Nilsson2017}  We treat the Ti-$3d$ submanifold with isolated bands of predominant $t_{2g}$ character as the strongly correlated subsystem, while higher-energy bands enter the description via the dynamical and spatial structure of the constrained random phase approximation (cRPA) \cite{Aryasetiawan2004} interaction and the $G^0W^0$ fermionic propagators. Technical details on the $GW$+DMFT implementation can be found in the End Matter.

All calculations in this study are performed at an inverse temperature of $\beta = 30$~eV$^{-1}$ ($T=387$~K) with a high-frequency cutoff of 200~eV, using a continuous-time quantum Monte Carlo impurity solver capable of treating dynamical screening, but limited to density-density interactions.\cite{Werner2006,Werner2010} The relatively high temperature is imposed by the memory requirements of the finite temperature $GW$ calculation; however, the main quantities of interest, such as the quasiparticle renormalization, orbital polarization, and the overall structure of the low-energy spectral weight, are governed by energy scales substantially larger than $T$, and thus remain essentially unaffected as the temperature is further reduced. As a result, the trends extracted from our data are expected to reflect the low-temperature behavior. To determine the superconducting critical temperature, we compute the electronic structure at a temperature higher than the expected $T_c$ but still well below the characteristic electronic energy scales.

\begin{figure}[t]
\begin{center}
\includegraphics[width=0.45\textwidth,angle=0]{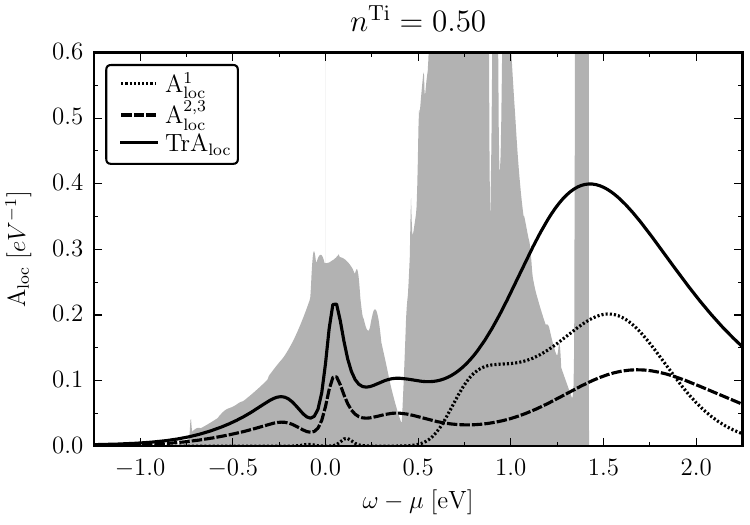}
\caption{Local $GW$+EDMFT spectral function at the stoichiometric filling (solid line). The dashed and dotted lines show the local spectral functions in the CF basis with two-fold degenerate occupied orbitals ($A^{2,3}$) and an almost empty orbital ($A^{1}$). The shaded region indicates the DFT DoS.} \label{fig:Aloc}
\end{center}
\end{figure}

Figure~\ref{fig:Aloc} shows the converged local spectral function of LiTi$_2$O$_4$ at the stoichiometric filling of 0.5 electrons per Ti site. The occupation of the well-localized (spread~$\sim$1.63 \AA$^2$) Wannier orbitals is degenerate, with 0.083 electrons per spin-orbital, corresponding to the spectrum shown by the full black line. 
The local interaction matrix ($\mathcal{U}$) for the Ti sites is calculated in the $GW$+EDMFT self-consistency loop for this degenerate Wannier basis, and then rotated to the diagonal crystal field (CF) basis for the solution of the impurity problem. This local rotation lifts the degeneracy, causing one of the three impurity levels to become nearly empty. The interaction parameters, on the other hand, remain the same (see End Matter). Hence, in the CF basis, we have two degenerate orbitals with approximately 0.445 electrons each, and one containing the remaining 0.055 electrons. The corresponding spectra are indicated by the dashed and dotted lines in Fig.~\ref{fig:Aloc}, respectively.

Inspired by the ionic gating experiments,\cite{Wei2021} we investigate the doping dependence of the system by tuning the chemical potential of the model. A fully consistent treatment would require performing separate cRPA and $G^0W^0$ (and later phonon) calculations for each target filling. This is an extremely demanding task, both computationally and conceptually, due to the difficulty of doping the band structure without altering the chemical constituents of the material. In the present work, we hence use the downfolded model for filling $n^\text{Ti}=0.5$ and scan the chemical potential to reach the desired occupation. This approximation is justified by the fact that the electronic degrees of freedom excluded from the low-energy model are sufficiently separated in energy from the correlated manifold.

\begin{figure}[tp!]
\begin{center}
\includegraphics[width=0.47\textwidth,angle=0]{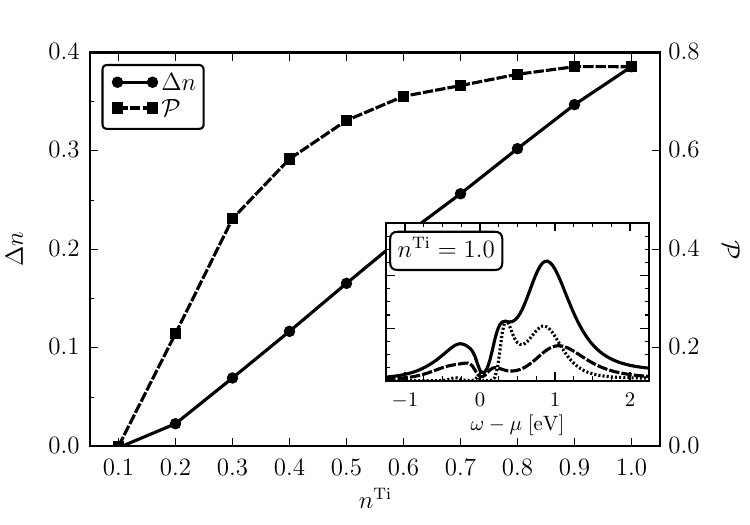}
\includegraphics[width=0.47\textwidth,angle=0]{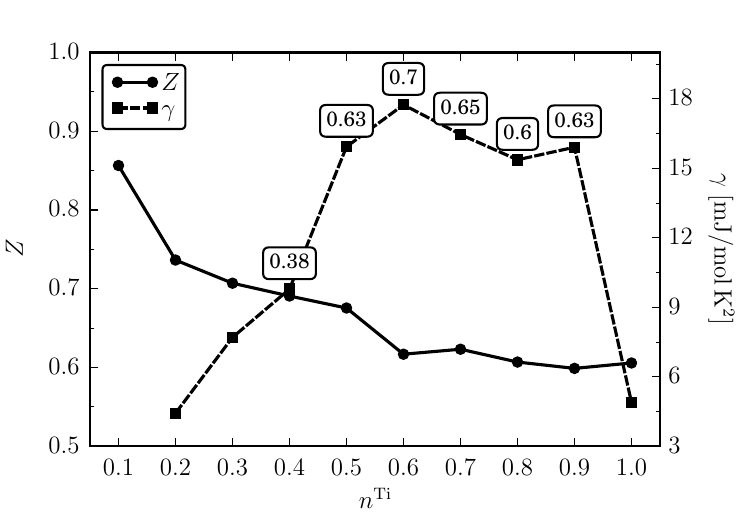}
\caption{Electronic properties as a function of the total filling of the Ti-$d$ shell, $n^\text{Ti}$. Top panel: density imbalance $\Delta n$ between the singly and doubly degenerate local levels (circles) and impurity polarization $\mathcal{P}$ (squares). The inset reports the insulating local spectral function at the $d^1$ configuration. Bottom panel: $Z$ factor (circles) and the Sommerfeld coefficient (squares) as a function of $n^\text{Ti}$. The labels indicate the value of the local interacting spectral function at the Fermi level (states eV$^{-1}$ atom$^{-1}$).} \label{fig:results_es}
\end{center}
\end{figure}

Insights into the evolution of the system at different fillings can be extracted from the diagonal elements of the CF impurity density matrix, $\text{diag}(\rho)\equiv (n, n+\Delta n, n+\Delta n)$, which satisfy the constraint $\mathrm{Tr}\rho=n^{\text{Ti}}$. From the values of $\Delta n$, indicated by the full black line in the top panel of Fig.~\ref{fig:results_es}, it follows that, upon approaching the $d^1$ configuration, the isolated level becomes depleted, whereas the two degenerate orbitals almost reach a half-filled configuration. The near-linear slope of the $\Delta n$ curve implies that the impurity orbital polarization that we define as $\mathcal{P}=2\Delta n/n^{\text{Ti}}$ almost saturates as $n^{\text{Ti}}\sim1$, as shown by the dashed line. At the $d^1$ occupation, a gap opens at the Fermi level, as a result of the almost complete depletion of the isolated orbital and the (near-)localization of the remaining electrons, see inset in the top panel of Fig.~\ref{fig:results_es}. 

In DMFT studies of metals the strength of the $e$–$e$ interactions is often quantified by the quasiparticle renormalization factor extracted from the local self-energy,
\begin{equation}
    Z=\left(1-\left.\frac{\partial\Im\Sigma^{\text{imp}}_{t_{2g}}\left(\omega_{n}\right)}{\partial\omega_{n}}\right|_{\omega_n\rightarrow 0}\right)^{-1},
\end{equation}
which is reported by black circles in the bottom panel of Fig.~\ref{fig:results_es}, together with the Sommerfeld coefficient
\begin{equation}
    \gamma=\frac{K_B^{2}\pi^{2}}{3} \left(1+\lambda \right) \frac{N_{\varepsilon_F}^{GW+\text{EDMFT}} }{Z}.
\end{equation}
In this expression, we used the $e$–$ph$ coupling $\lambda=0.54$ derived from the phonon calculation described below and in the End Matter.  We furthermore evaluated the height of the quasiparticle peak using the approximate formula $N_{\varepsilon_F}^{GW+\text{EDMFT}}=\beta G_{\mathrm{loc}}\left(\tau=\beta/2 \right)/\mathrm{atom}$,\cite{Gull2008} see the labels next to the squares in Fig.~\ref{fig:results_es}.  
Due to the lack of particle-hole symmetry, $\Im\Sigma^{\text{imp}}_{t_{2g}}$ is not diverging even in the insulating regime, resulting in $Z>0$ for $n^\text{Ti}=1$. Also, since the DoS does not vanish completely in our $\beta=30$ eV$^{-1}$ calculation, the $\gamma$ coefficient remains nonzero in this limit. 

\begin{table}[tb!]
  \centering
  \caption{Comparison between the experimental measurements on LTO \cite{Sun2004,Tang2006} and our theoretical results. The numerical data are reported for the fillings between 0.6 and 0.8 corresponding to the highest $T_c$.}
  \label{tab:ExperimentsComparison}
  \small
  \begin{tabular}{@{}crr@{}}
    \toprule
     \hline
    \midrule
     \hline
     Parameters                              & Ref.~\onlinecite{Sun2004} & Present work \\
     \hline
    \midrule
     \hline
     $T_c$ (K)                               & $11.4\pm0.3$              & 10.6$\pm$1.5    \\
     \hline
     $\Delta$ (meV)                          &  1.97                     & 1.82$\pm$0.37   \\
     \hline
     $Z$                                     & $\sim0.65$                & 0.62$\pm$0.01  \\
     \hline
     $\gamma$ (mJ mol$^{-1}$ K$^{-2}$)       & $19.15\pm0.20$            & 16.7$\pm$1.2   \\
     \hline
     $N_{\varepsilon_F}$ (states eV$^{-1}$ atom$^{-1}$) & $0.70\pm0.01$  & 0.65$\pm$0.05  \\
     \hline
     $2\Delta/k_B T_c$                       & $\sim4.0$                 & $4.0\pm0.3$     \\
     \hline
     $\lambda$                               & 0.65                      & 0.54 \\
     \hline
    \bottomrule
  \end{tabular}
\end{table}

In the occupation range corresponding to the highest $T_c$, our calculations yield almost quantitative agreement with the available experimental benchmarks, see Tab.~\ref{tab:ExperimentsComparison}. In particular, the Sommerfeld coefficient extracted from our $GW$+EDMFT solution reproduces the most recent reported value of $\gamma = 19.15$~mJ~mol$^{-1}$~K$^{-2}$ with good accuracy for a wide range of Ti occupations. Likewise, the quasiparticle renormalizations $Z=0.6$–$0.75$ fall within the range inferred from photoemission and transport analyses, and they are consistent with the  expected moderate $e$–$e$ correlations. Finally, the emergence of insulating states flanking the $d^0$–$d^1$ region is fully consistent with the experimentally observed suppression of metallicity for both electron depletion and electron accumulation in the ionic gating experiments.

\begin{figure}[htp!]
\includegraphics[width=0.45\textwidth,angle=0]{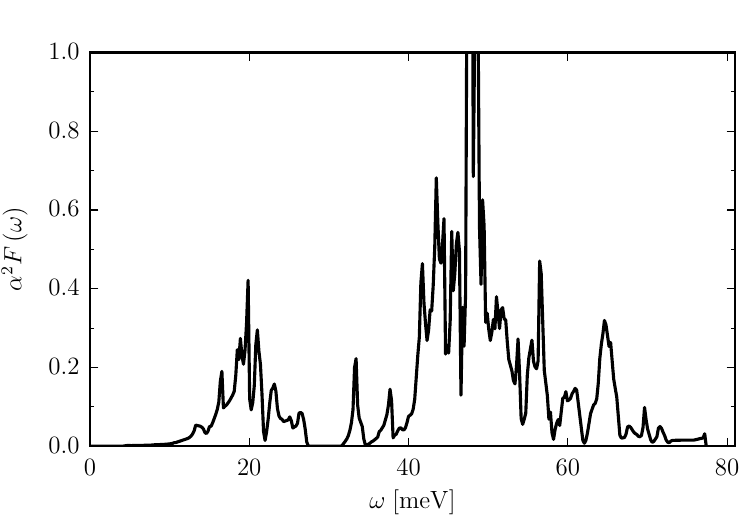}
\includegraphics[width=0.45\textwidth,angle=0]{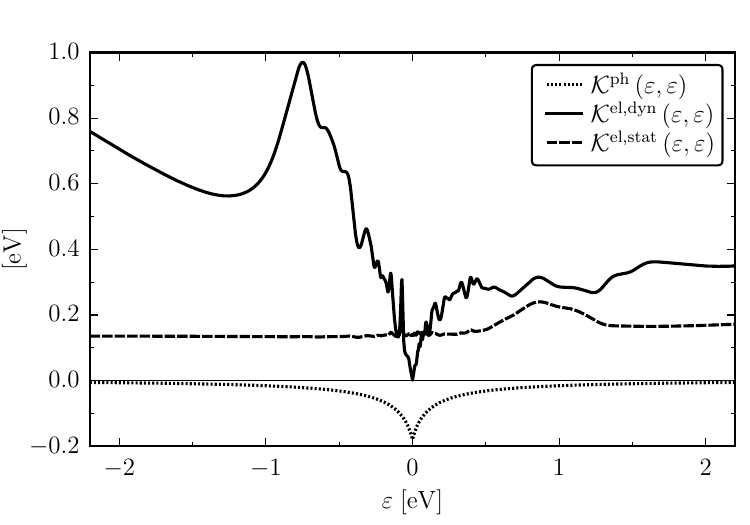}
\caption{Upper panel: Eliashberg function $\alpha^2F(\omega)$ of LTO. Bottom panel: diagonal components of the contributions to the SCDFT Kernel $\mathcal{K}$ computed at $T=0.05$~K.} \label{fig:Kernels}
\end{figure}

{\it Superconductivity --} To calculate the superconducting critical temperature $T_c$, we combine our $GW$+EDMFT results with density functional theory for superconductors (SCDFT).\cite{Oliveira1988,Luders2005,Marques2005} SCDFT provides a first-principles framework to determine $T_c$, based on a gap equation in which the pairing interaction and mass renormalization are encoded in two kernels, $\mathcal{Z}^{\mathrm{ph}}_{n\mathbf{k}}$ and $\mathcal{K}_{n\mathbf{k},n'\mathbf{k}'}$, which are, respectively, diagonal and off-diagonal in the Kohn-Sham eigenvalue index. We detail our implementation of the formalism in the End Matter. The superconducting order parameter $\Delta_{n\mathbf{k}}$ is obtained as the self-consistent solution of the gap equation
\begin{equation}
\Delta_{n\mathbf{k}}=-\mathcal{Z}^{\mathrm{ph}}_{n\mathbf{k}}\,\Delta_{n\mathbf{k}}-\frac{1}{2}\sum_{n'\mathbf{k}'}\mathcal{K}_{n\mathbf{k},n'\mathbf{k}'}\frac{\tanh\!\big[(\beta/2)E_{n'\mathbf{k}'}\big]}{E_{n'\mathbf{k}'}}\Delta_{n'\mathbf{k}'},
\label{eq:gap-k}
\end{equation}
where $\beta$ is the inverse temperature and \(E_{n\mathbf{k}}=\sqrt{\varepsilon_{n\mathbf{k}}^2+\Delta_{n\mathbf{k}}^2}\) (with $\varepsilon_{n\mathbf{k}}$ the Kohn-Sham eigenvalues) denotes the quasiparticle energies. In the following, the diagonal kernel $\mathcal{Z}^\text{ph}_{n\mathbf{k}}$ is assumed to contain only the $e$–$ph$ contribution, while the pairing kernel is decomposed into phononic and electronic parts $\mathcal{K}_{n\mathbf{k},n'\mathbf{k}'}^{\mathrm{ph}}$ and $\mathcal{K}_{n\mathbf{k},n'\mathbf{k}'}^{\mathrm{el}}$.  

We report in the upper panel of Fig.~\ref{fig:Kernels} the Eliashberg function $\alpha^2F\left(\omega\right)$, which is needed to define the phononic kernels (Eqs.~\eqref{eq:Zph_en} and \eqref{eq:Kph_en} in the End Matter), and from which we can also estimate the magnitude of the $e$–$ph$ coupling used in this study: $\lambda = 2\int \frac{d\omega}{\omega}\alpha^2 F(\omega)=0.54$, in agreement with previous results.\cite{Oda1994,Oda1996}

The electronic Kernel $\mathcal{K}_{n\mathbf{k},n'\mathbf{k}'}^{\mathrm{el}}$ can be separated into a static and a dynamic contribution, so that\cite{Akashi2013,Akashi2014}
\begin{equation}
\mathcal{K}_{n\mathbf{k},n'\mathbf{k}'}=\mathcal{K}_{n\mathbf{k},n'\mathbf{k}'}^{\mathrm{ph}}+\mathcal{K}_{n\mathbf{k},n'\mathbf{k}'}^{\mathrm{el,stat}}+\mathcal{K}_{n\mathbf{k},n'\mathbf{k}'}^{\mathrm{el,dyn}}.
\label{eq:Kernels}
\end{equation}
The static electronic pairing kernel $\mathcal{K}_{n\mathbf{k},n'\mathbf{k}'}^{\mathrm{el,stat}}$ is calculated from the static limit of the screened interaction 
$W(i\Omega_n)$ as
\begin{equation}
\mathcal{K}_{n\mathbf{k},n'\mathbf{k}'}^{\mathrm{el,stat}}=W_{n\mathbf{k},n'\mathbf{k}'}\left(i\Omega_n=0\right),
\label{eq:Kel-stat}
\end{equation}
whereas the dynamic component can be expressed in terms of the anomalous Green's functions $F_{n\mathbf{k}}(i\omega)=\frac{2E_{n\mathbf{k}}}{\omega^2+E^2_{n\mathbf{k}}}$ evaluated at the DFT level and the frequency-dependent part of the screened interaction $W^{\mathrm{dyn}}\left(i\Omega_n\right)=W\left(i\Omega_n\right)-W\left(0\right)$ as
\begin{align}
\mathcal{K}_{n\mathbf{k},\,n'\mathbf{k}'}^{\mathrm{el},\,\mathrm{dyn}}=&\lim_{\Delta_{n\mathbf{k}}\to0}\frac{1}{\beta^{2}}\frac{1}{\tanh(\beta E_{n\mathbf{k}}/2)}\frac{1}{\tanh(\beta E_{n'\mathbf{k}'}/2)} \nonumber \\&\sum_{\omega_{1}\omega_{2}}F_{n\mathbf{k}}(i\omega_{1})F_{n'\mathbf{k}'}(i\omega_{2})W_{n\mathbf{k},n'\mathbf{k}'}^{\mathrm{dyn}}(i\omega_{1}-i\omega_{2}).
\label{eq:Kel-dyn}
\end{align}

We use the momentum-dependent screened interaction obtained from the converged $GW$+EDMFT solution, which accounts for both local (EDMFT) and long-range ($GW$) correlations and dynamical screening. The bosonic propagator is usually calculated from numerically less advanced schemes -- such as RPA,\cite{Akashi2013,Errea2020,Sanna2020,Christiansson2024} the adiabatic local density approximation,\cite{Tsutsumi2020} or plain $GW$\cite{Christiansson2022} -- and used in the kernels above. The novel aspect of the present work is the application (within SCDFT) of a non-perturbative many-body method to a compound characterized by significant local electronic correlations. 

The static and dynamic electronic contributions to the non-diagonal kernel are plotted in the bottom panel of Fig.~\ref{fig:Kernels} with thick and dashed lines, respectively. The static contribution is, as expected, nearly constant, whereas the dynamic component displays a pronounced energy dependence and as a consequence allows for an enhancement of the retardation effect that reduces the effective Coulomb repulsion,\cite{Akashi2013,Akashi2014} albeit with features that are somewhat broadened compared to those obtained using the non-interacting DoS.\footnote{See for instance Fig.4 in Ref.~\onlinecite{Akashi2013}}

\begin{figure}[tp!]
\includegraphics[width=0.45\textwidth,angle=0]{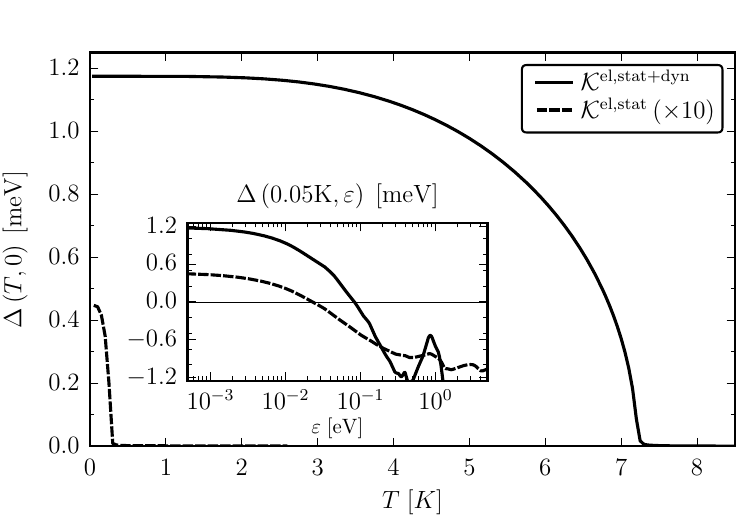}
\caption{Gap amplitude $\Delta(T,0)$ of stoichiometric LTO as a function of temperature (solid line), yielding $T_c=7.2$ K. The dashed line shows the result obtained with only the static interaction, which clearly underestimates $T_c$. Inset: energy structure of the low temperature BCS order parameter $\Delta(T=0.05~\mathrm{K},\varepsilon)$. The structure near $\varepsilon=0.8$ is likely of plasmonic origin, since it is not present in the calculation with static kernel. All dashed lines are $\times 10$ magnified for clarity.} \label{fig:Delta}
\end{figure}

In practice, we reformulate the gap equation \eqref{eq:gap-k} in terms of integrals over the density of states (Eq.~\eqref{eq:gap-energy} in the End Matter). For each doping level of the electronic structure, we then solve this equation at a given temperature. Starting below the expected $T_c$, the self-consistent solution converges to a stable energy-dependent order parameter $\Delta(T,\varepsilon)$ that balances the repulsive and attractive interactions encoded in the kernels. A nonvanishing order parameter at the Fermi level, $\Delta(T,0)$, which represents the gap amplitude, signals the existence of a superconducting solution. By analyzing the evolution of $\Delta(T,0)$, with kernels recalculated at progressively increasing $T$, the critical temperature can be identified by the vanishing of the the gap amplitude, $\Delta(T_c,0)=0$. The order parameter at $T=0.02$~K, associated with LTO at the stoichiometric filling, is reported in the inset of Fig.~\ref{fig:Delta}. The nonzero value for $\varepsilon\rightarrow 0$ demonstrates the existence of a superconducting solution, and we find that superconductivity persists up to approximately 7.2~K (main panel of Fig.~\ref{fig:Delta}), close to the first reported experimental value of 11.5 K.\cite{Johnston1973} An important  observation is that the superconducting state becomes extremely weak when only the static part of the $e$–$e$ kernel is considered (dashed line in Fig.~\ref{fig:Delta}), with a substantially smaller order parameter and much lower $T_c$,  compared to the typical reduction by a factor of 2-5.\cite{Akashi2014,Christiansson2022,Christiansson2024} This demonstrates that the dynamical structure of the interaction plays a central role in the pairing mechanism. When the retardation effect is omitted, a significant negative contribution to the integral of $\mathcal{K}\Delta$ is missing in the gap equation, which suppresses the superconducting order. Only with the full dynamical kernel a $T_c$ of the correct order of magnitude is obtained. Within the $GW$+EDMFT framework, the collective charge fluctuations (plasmons) captured by the bosonic self-consistency loop are responsible for the retardation effect. This demonstrates that plasmons are essential to explain the remarkably high $T_c$ of LTO. 

\begin{figure}[t]
\includegraphics[clip,width=0.47\textwidth,angle=0]{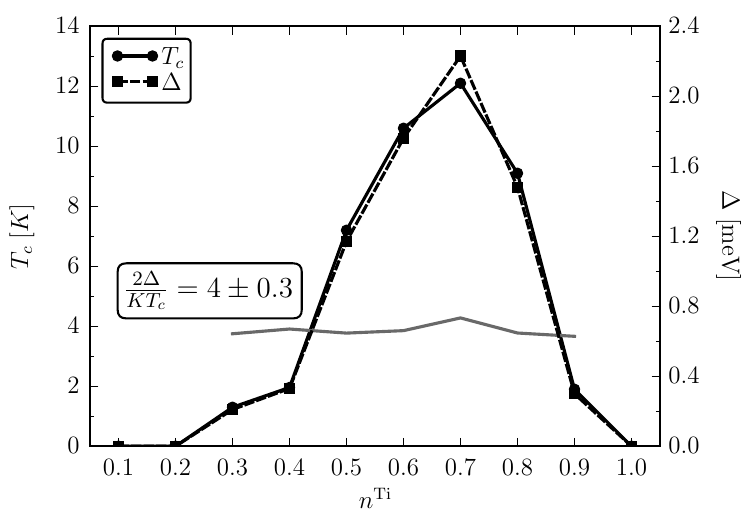}
\caption{Superconducting dome of LiTi$_2$O$_4$. The thick line indicates the critical temperature, while the dashed line shows the BCS order parameter. The gray horizontal line represents the BCS ratio for the $d$ shell occupations with a superconducting solution.} \label{fig:Dome}
\end{figure}

Figure~\ref{fig:Dome} summarizes the superconducting properties across the entire doping range and reveals a dome-shaped $T_c$ in good agreement with ionic gating measurements. In the metallic region between the $d^0$ and $d^1$ configurations, the superconducting solution emerges below $12.3$~K, which is very close to the maximum experimental $T_c$ (Tab.~\ref{tab:ExperimentsComparison}). The figure reports both the critical temperature and the zero-temperature gap amplitude.  Also the maximum gap of approximately 2.2 meV is in good agreement with experiment. The ratio of gap amplitde and $T_c$ is compatible with the BCS expectation over the entire doping range, which confirms the internal consistency of the calculation.

{\it Conclusions --} We used $GW$+EDMFT to calculate the electronic structure and screened interaction of LiTi$_2$O$_4$, and employed the dynamically screened $W(\omega)$ in the electronic kernel of SCDFT to estimate the superconducting $T_c$. This computational framework yields excellent agreement with experiment for the metal-insulator transitions, the mass enhancement, Sommerfeld coefficient, and filling-dependent $T_c$, while SCDFT calculations performed with the static screened interaction $W(\omega=0)$ underestimate $T_c$ by more than a factor of 10, proving the relevance of the frequency structure of the $e$-$e$ interaction. The coupling to plasmons, encoded in the frequency dependence of $W$, hence plays an important role in the pairing, and explains the remarkably high superconducting $T_c$ in LiTi$_2$O$_4$, which should be characterized as a plasmon-assisted superconductor. More generally, the accurate prediction of thermodynamic, spectral, and transport-related quantities across the valence window accessed in ionic-gating experiments suggests that the presented combination of the $GW$+EDMFT and SCDFT frameworks captures the essential physics of moderately correlated superconductors. It would be interesting to apply it to other compounds with potentially significant plasmonic contribution to the pairing, such as layered $\beta$-nitrides\cite{Akashi2012} or BiS$_2$-based superconductors.\cite{Morice2017}

{\it Acknowledgements --} The calculations have been performed on the beo05 cluster at the University of Fribourg. PW acknowledges support from ERC Consolidator Grant No.~724103 and the Swiss National Science Foundation via NCCR Marvel. RA is supported by Grant-in-Aid for Scientific Research from JSPS, KAKENHI Grant  No. 25H01246, No. 25H01252, No. 24H00190, JST K-Program JPMJKP25Z7, RIKEN TRIP initiative (RIKEN Quantum, Advanced General Intelligence for Science Program, Many-body Electron Systems).

\begin{center}
\textbf{END MATTER}
\end{center}

\begin{center}
\textbf{$GW$+EDMFT calculation}
\end{center}
Our {\it ab-initio} calculation starts from the DFT electronic structure obtained using the full-potential linearized augmented plane-wave code FLEUR \footnote{The FLEUR group, The FLEUR project, www.flapw.de.} with the GGA functional\cite{Perdew1996} and a $18\times18\times18$ $\mathbf{k}$-grid, followed by the construction of twelve maximally localized $t_{2g}$-like Wannier functions centered on the four Ti sites in the primitive unit cell via the Wannier90 library.\cite{Wannier90} The cRPA and single-shot $G^0W^0$ calculations, which define the bare bosonic and fermionic propagators  for the $GW$+EDMFT self-consistency loop, were performed using the SPEX code\cite{Friedrich2010} with an $8\times8\times8$ $\mathbf{k}$-grid. Unoccupied states up to $\sim$50~eV were included in the calculation of the polarization function and self-energy.

In the Wannier representation, the local tight-binding bare fermionic Hamiltonian $H_{t_{2g}}\left(\mathbf{R}=0\right)$ contains four sites per unit cell, with three orbitals each. While the diagonal levels on each site are degenerate, blocks associated with different sites have local off-diagonal hopping terms which differ by a phase factor between the different sites. Within EDMFT,\cite{Sun2002} the lattice model is mapped onto four impurity problems, which become locally equivalent after applying the corresponding local SO(3) rotations that diagonalize the $H_{t_{2g}}\left(\mathbf{R}=0\right)$ block of each titanium site, reducing the problem to the numerical solution of a single impurity in the so-called crystal field (CF) basis.\cite{Petocchi2021} The resulting impurity self-energy $\Sigma^{\text{imp}}$ and polarization $\Pi^{\text{imp}}$ are then rotated back to the Wannier basis, preserving the site-dependent phase modulations of the inter-orbital components. 

As a consequence of the orbital degeneracy, the interaction obtained from the cRPA calculation is locally rotationally invariant when expressed in spin–orbital space, and therefore retains the same structure also in the crystal field basis. Its magnitude, $U=4.35$~eV with a Hund’s coupling $J=0.5$~eV, combined with a bandwidth of $\sim 2.5$~eV, indicates that the system should exhibit sizeable correlation effects, despite the low band filling.

\begin{center}
\textbf{SCDFT calculation}
\end{center}

\vspace{1em}

In density functional theory for superconductors (SCDFT) \cite{Oliveira1988,Luders2005,Marques2005} the superconducting order parameter $\Delta_{n\mathbf{k}}$ is obtained as the self-consistent solution of the gap equation
\begin{equation}
\Delta_{n\mathbf{k}}=-\mathcal{Z}^{\mathrm{ph}}_{n\mathbf{k}}\,\Delta_{n\mathbf{k}}-\frac{1}{2}\sum_{n'\mathbf{k}'}\mathcal{K}_{n\mathbf{k},n'\mathbf{k}'}\frac{\tanh\!\big[(\beta/2)E_{n'\mathbf{k}'}\big]}{E_{n'\mathbf{k}'}}\Delta_{n'\mathbf{k}'},
\label{eq:gap-k}
\end{equation}
where $\beta$ is the inverse temperature and \(E_{n\mathbf{k}}=\sqrt{\varepsilon_{n\mathbf{k}}^2+\Delta_{n\mathbf{k}}^2}\) (with $\varepsilon_{n\mathbf{k}}$ the Kohn-Sham eigenvalues) denotes the quasiparticle energies. In this study, the diagonal kernel $\mathcal{Z}^\text{ph}_{n\mathbf{k}}$ is assumed to contain only the $e$–$ph$ contribution, while the pairing kernel is decomposed into phononic and electronic parts $\mathcal{K}_{n\mathbf{k},n'\mathbf{k}'}^{\mathrm{ph}}$ and $\mathcal{K}_{n\mathbf{k},n'\mathbf{k}'}^{\mathrm{el}}$. We calculate the electron-phonon matrix elements $g_{\mu{\bf q}}^{n{\bf k},n'{\bf k'}}$ on a $6\times6\times6$ ${\bf q}$-grid and a $18\times18\times18$ ${\bf k}$-grid using density functional perturbation theory, \cite{Baroni2001} as implemented in \textsc{Quantum ESPRESSO}\cite{Giannozzi2009,Giannozzi2017} ($\mu$ and $n$ refer to the phonon mode and band index, respectively). The phononic contributions we consider are calculated using the Lüders-Marques functionals. \cite{Luders2005,Marques2005} These are of the form 
\begin{align}
\mathcal{Z}^{\textrm{ph}}_{n{\bf k}} =& \frac{1}{\tanh[(\beta/2)\varepsilon_{n{\bf k}}]}\sum_{n {\bf k'}}\sum_{\mu {\bf q}}\left| g^{n{\bf k},n'{\bf k'}}_{\mu{\bf q}} \right|^2\nonumber\\ 
&\times\left[ J(\varepsilon_{n{\bf k}},\varepsilon_{n'{\bf k'}},\omega_{\mu{\bf q}}) + J(\varepsilon_{n{\bf k}},-\varepsilon_{n'{\bf k'}},\omega_{\mu{\bf q}}) \right]
\label{eq:Zph}
\end{align}
and
\begin{align}
\mathcal{K}^{\textrm{ph}}_{n{\bf k}n'{\bf k'}} =& \frac{1}{\tanh[(\beta/2)\varepsilon_{n{\bf k}}]}\frac{1}{\tanh[(\beta/2)\varepsilon_{n'{\bf k'}}]}\sum_{\mu {\bf q}}\left| g^{n{\bf k},n'{\bf k'}}_{\mu{\bf q}} \right|^2\nonumber\\
& \times \left[ I(\varepsilon_{n{\bf k}},\varepsilon_{n'{\bf k'}},\omega_{\mu{\bf q}}) - I(\varepsilon_{n{\bf k}},-\varepsilon_{n'{\bf k'}},\omega_{\mu{\bf q}}) \right],
\label{eq:Kph}
\end{align}
with the functions $I$ and $J$ defined in terms of the Fermi-Dirac ($n_F(\varepsilon)$) and Bose-Einstein ($n_B(\omega)$) distributions:
\begin{align}
&I(\varepsilon,\varepsilon',\omega)= n_F(\varepsilon)n_F(\varepsilon')n_B(\omega)\nonumber\\
&\quad\times \left( \frac{e^{\beta \varepsilon}-e^{\beta(\varepsilon'+\omega)}}{\varepsilon-\varepsilon'-\omega} - \frac{e^{\beta \varepsilon'}-e^{\beta(\varepsilon+\omega)}}{\varepsilon-\varepsilon'+\omega} \right),\\
%
&J(\varepsilon,\varepsilon',\omega)= \tilde{J}(\varepsilon,\varepsilon',\omega)-\tilde{J}(\varepsilon,\varepsilon',-\omega),\\
&\tilde{J}(\varepsilon,\varepsilon',\omega)=-\frac{n_F(\varepsilon)+n_B(\omega)}{\varepsilon-\varepsilon'-\omega}\nonumber\\
&\quad\times \left( \frac{n_F(\varepsilon')-n_F(\varepsilon-\omega)}{\varepsilon-\varepsilon'-\omega} - \beta n_F(\varepsilon-\omega)n_F(-\varepsilon+\omega) \right).
\end{align}
An alternative form of the $e$–$ph$ kernels has been proposed by Sanna and co-workers,\cite{Sanna2020} showing an improvement over the original functional for a wide range of systems. However, since we are interested here in a pairing mechanism of dominant electronic origin, we choose the standard Lüders-Marques functional. 

The Eliashberg function
\begin{equation}
\alpha^{2}F\left(\omega\right)=\sum_{\mu\mathbf{q}}\frac{\delta\left(\omega-\omega_{\mu \mathbf{q}}\right)}{N_{\varepsilon_{F}}^{DFT}}\sum_{n\mathbf{k},n'\mathbf{k}'}\left|g_{\mu\mathbf{q}}^{n\mathbf{k},n'\mathbf{k}'}\right|^{2}\delta\left(\varepsilon_{n\mathbf{k}}\right)\delta\left(\varepsilon_{n'\mathbf{k}'}\right)
\label{eq:Eliashberg}
\end{equation}
encodes the $e$–$ph$ coupling constants $g_{\mu,{\bf q}}^{n{\bf k},n'{\bf k'}}$ and related energies $\omega_{\mu {\bf q}}$ in a single frequency-dependent function evaluated at the Fermi level of the DFT bandstructure. This allows to re-express the phononic kernels of Eq.~\eqref{eq:Zph} and Eq.~\eqref{eq:Kph} in an energy-averaged formulation as
\begin{align}
&\mathcal{Z}^\text{ph}\left(\varepsilon\right)=-\frac{1}{\tanh(\beta/2\varepsilon)}\int d \varepsilon' \nonumber\\
& \quad\times\int d\omega \alpha^2F(\omega) \left[ J(\varepsilon,\varepsilon',\omega)+J(\varepsilon,-\varepsilon',\omega)\right],\label{eq:Zph_en}\\
\nonumber \\
&\mathcal{K}^\text{ph}\left(\varepsilon,\varepsilon'\right)=\frac{2}{\tanh(\varepsilon\beta/2)\tanh(\varepsilon'\beta/2)}\frac{1}{N_{\varepsilon_{F}}^{DFT}} \nonumber\\
&\quad \times \int d\omega \alpha^2F(\omega) \left[ I(\varepsilon,\varepsilon',\omega)-I(\varepsilon,-\varepsilon',\omega)\right] \label{eq:Kph_en}.
\end{align}
Furthermore, it follows that the typical dimensionless $e$–$ph$ coupling constant $\lambda$, is related to the kernels as \cite{Luders2005,Marques2005}
\begin{equation}
\lambda =0.54 \approx \mathcal{Z}^\text{ph}(0)\approx-\mathcal{K}^\text{ph}\left(0,0\right)N_{\varepsilon_{F}}^{DFT}.
\end{equation}

The electronic Kernel $\mathcal{K}_{n\mathbf{k},n'\mathbf{k}'}^{\mathrm{el}}$ can further be separated into a static and a dynamic contribution, so that\cite{Akashi2013,Akashi2014}
\begin{equation}
\mathcal{K}_{n\mathbf{k},n'\mathbf{k}'}=\mathcal{K}_{n\mathbf{k},n'\mathbf{k}'}^{\mathrm{ph}}+\mathcal{K}_{n\mathbf{k},n'\mathbf{k}'}^{\mathrm{el,stat}}+\mathcal{K}_{n\mathbf{k},n'\mathbf{k}'}^{\mathrm{el,dyn}}.
\label{eq:Kernels}
\end{equation}
The static electronic pairing kernel $\mathcal{K}_{n\mathbf{k},n'\mathbf{k}'}^{\mathrm{el,stat}}$ is calculated from the static limit of the screened interaction (or bosonic propagator) $W(i\Omega_n)$ as
\begin{equation}
\mathcal{K}_{n\mathbf{k},n'\mathbf{k}'}^{\mathrm{el,stat}}=W_{n\mathbf{k},n'\mathbf{k}'}\left(i\Omega_n=0\right),
\label{eq:Kel-stat}
\end{equation}
whereas the dynamic component can be expressed in terms of the anomalous Green's functions $F_{n\mathbf{k}}(i\omega)=\frac{2E_{n\mathbf{k}}}{\omega^2+E^2_{n\mathbf{k}}}$ evaluated at the DFT level and the frequency-dependent part of the screened interaction $W^{\mathrm{dyn}}\left(i\Omega_n\right)=W\left(i\Omega_n\right)-W\left(0\right)$ as
\begin{align}
\mathcal{K}_{n\mathbf{k},\,n'\mathbf{k}'}^{\mathrm{el},\,\mathrm{dyn}}=&\lim_{\Delta_{n\mathbf{k}}\to0}\frac{1}{\beta^{2}}\frac{1}{\tanh(\beta E_{n\mathbf{k}}/2)}\frac{1}{\tanh(\beta E_{n'\mathbf{k}'}/2)} \nonumber \\&\sum_{\omega_{1}\omega_{2}}F_{n\mathbf{k}}(i\omega_{1})F_{n'\mathbf{k}'}(i\omega_{2})W_{n\mathbf{k},n'\mathbf{k}'}^{\mathrm{dyn}}(i\omega_{1}-i\omega_{2}).
\label{eq:Kel-dyn}
\end{align}

In our implementation, we use the momentum-dependent screened interaction obtained from the converged $GW$+EDMFT solution, which accounts for both local (EDMFT) and long-range ($GW$) correlations and dynamical screening. The relevant energy scales are very different for the $e$–$e$ and $e$–$ph$ contributions to the kernels, with the latter being relevant only within a window on the order of the Debye frequency.\cite{Marques2005} To resolve this, a very fine sampling of the $\mathbf{k}$-mesh is required to accurately capture the contribution from states around the Fermi level, which renders a fully self-consistent $GW$+EDMFT solution computationally prohibitive. To overcome this limitation, it is advantageous to express both the kernels and the order parameter in the energy domain,\cite{Marques2005,Kawamura2017} thereby avoiding the need for ultra-dense momentum sampling for $W_{n\mathbf{k},n'\mathbf{k}'}(i\Omega_n)$. We refer the reader to Refs.~\onlinecite{Akashi2013,Kawamura2017,Christiansson2022,Christiansson2024} for a detailed description of the formalism with a dynamically screened $e$–$e$ interaction. At the cost of losing information on the symmetry of the order parameter, which becomes $s$-wave by construction, the gap equation is then recast into the form
\begin{align}
\Delta(T,\varepsilon)=&-\mathcal{Z}^{ph}(\varepsilon)\Delta(\varepsilon)\nonumber \\&-\frac{1}{2}\int d\varepsilon'N(\varepsilon')\mathcal{K}(\varepsilon,\varepsilon')\frac{\tanh\big[(\beta/2)E'\big]}{E'}\Delta(T,\varepsilon'),
\label{eq:gap-energy}
\end{align}
where $N(\varepsilon)$ is the DoS. In the context of this study it is important to emphasize that the energy representation of the $e$–$e$ kernels is defined by numerical averages over iso-energetic surfaces,
\begin{equation}
\mathcal{K}(\varepsilon,\varepsilon')=\sum_{n\mathbf{k},n'\mathbf{k}'}\frac{\delta(\varepsilon-\varepsilon_{n\mathbf{k}})}{N(\varepsilon)}\frac{\delta(\varepsilon'-\varepsilon_{n'\mathbf{k}'})}{N(\varepsilon')}\mathcal{K}_{n\mathbf{k},n'\mathbf{k}'},
\label{eq:Kel-energy}
\end{equation}
which, in the original formulation, was carried out using the non-interacting DoS. Relying on the latter has the drawback that its profile is rigid under changes in the filling, i.e. the density dependence of the electronic kernel before performing the $\mathbf{k}$-space integration is not fully captured in the energy representation.  In principle, this Hilbert-transform mapping can be performed with any suitably defined spectral function, and we hence tried to replace the DFT DoS with the fully interacting spectral function evaluated at each $\mathbf{k}$ point. However, the numerical inaccuracies associated with analytic continuation prevented a stable convergence of the gap equation \eqref{eq:gap-energy}. A practical compromise, which transfers a substantial portion of the doping dependence onto the energy axis while avoiding analytic continuation errors, is achieved by employing the $G^0W^0$ spectral function, corresponding to the bare fermionic propagator from the downfolding,\cite{Christiansson2022} which is evaluated directly on the real-frequency axis.

\bibliography{bibliography}

\end{document}